\documentclass[twocolumn,preprintnumbers,amsmath,amssymb,prb]{revtex4}% Physical Review B
\usepackage[english]{babel}
\usepackage{graphicx}% Include figure files
\usepackage{dcolumn}% Align table columns on decimal point
\usepackage{bm}% bold math
\usepackage[version=3]{mhchem}
\begin{document}

\preprint{Published in Journal of Physical Chemistry C 120 (2016) 2829--2836. DOI: 10.1021/acs.jpcc.5b11794}

\title{First-Principle Study of Adsorption and Desorption of Chlorine on
Cu(111) Surface. \\ Does Chlorine or Copper Chloride Desorb?}

\author{Tatiana V. Pavlova$^{1}$}
\email{pavlova@kapella.gpi.ru}
\author{Boris V. Andryushechkin$^{1}$}
\author{Georgy M. Zhidomirov$^{1,2}$}

\affiliation{$^{1}$Prokhorov General Physics Institute of the Russian Academy of Sciences, Moscow, Russia}
\affiliation{$^{2}$G.K.Boreskov Institute of Catalysis, Siberian Branch
of the Russian Academy of Sciences, Novosibirsk, Russia}

%\date{\today}% It is always \today, today,
             %  but any date may be explicitly specified

\begin{abstract}
First-principle density-functional calculations have been applied to
study the interaction of molecular chlorine with  the (111) plane of
copper. Using transition-state search method, we considered the
elementary processes (Cl$_2$ dissociation, adsorption, diffusion,
association and desorption) on the chlorinated Cu(111) surface. A
systematic study of possible desorption pathways has been carried
out for different species (Cl, Cl$_2$, CuCl, CuCl$_2$, and Cu) at
various chlorine coverage. As a result, we concluded that chlorine
monolayer irrespective of the coverage desorbs in the form of CuCl
molecules from step edges.

\end{abstract}

\maketitle

\section{Introduction}
The adsorption of chlorine on Cu(111) has been a subject of
extensive experimental studies since the 1970s
\cite{Goddard-Lambert,Goldmann1,Goldmann2,Jones89,Eltsov91,Sakurai93,EuroPhysLett86,
SS87,PRL87,SS88,Jones96,Way,AndryushechkinSS00,AndryushechkinUFN00,Shard04,Cu-Cl-Chains}.
In the first work, Goddard and Lambert \cite{Goddard-Lambert}
studied the Cl/Cu(111) system with temperature  programmed
desorption (TPD), Auger electron spectroscopy (AES), work function
measurements and low-energy electron diffraction (LEED). According
to TPD and AES spectra obtained in Ref. \cite{Goddard-Lambert}, only
one chemical state (chemisorbed chlorine) has been identified on the
Cu(111) surface. In addition, TPD experiments showed that chlorine
atoms were the only desorption product \cite{Goddard-Lambert}.
Eltsov et al. \cite{Eltsov91} considered the Cl$_{2}$ adsorption on
Cu(111) in the temperature range of 150-300~K and detected two
chemical states on the chlorinated Cu(111) surface: chemisorbed
chlorine and copper (I) chloride. Both states were clearly visible
in TPD and AES spectra. Analysis of desorption products reveals that
the chloride phase desorbs in the form of Cu$_{3}$Cl$_{3}$ trimers,
whereas chemisorbed chlorine desorbs as CuCl molecules
\cite{Eltsov91}. Formation of two chemical states has been also
confirmed in the work by Walter et al. \cite{Jones96}.

First structural studies of chemisorbed chlorine phases on Cu(111)
have been performed with LEED
\cite{Goddard-Lambert,Jones89,Jones96}. Goddard and Lambert \cite
{Goddard-Lambert} found that the
($\sqrt{3}\times\sqrt{3}$)R30$^\circ$  structure developed at
chlorine coverage of 0.33 ML is replaced by several complex
diffraction patterns containing the 6-spot triangles around the
$\sqrt{3}$ positions at higher coverages. Similar diffraction
patterns were also reported in Refs. \cite{Jones89,Jones96}. These
patterns have been explained in Refs.
\cite{Goddard-Lambert,Jones89,Jones96} by diffraction on the
uniformly compressed hexagonal lattices of chlorine  rotated by 30
degrees with respect to basic directions of Cu(111). Note that in
the early scanning tunneling microscopy (STM) study this model has
been confirmed \cite{Sakurai93}. However in 2000, the model of the
uniform compression was revised in the STM works by
Andryushechkin et al. \cite{AndryushechkinSS00,AndryushechkinUFN00}.
Compression of the chlorine lattice  on Cu(111) was found to be
uniaxial and non-uniform. According to
Refs.\cite{AndryushechkinSS00,AndryushechkinUFN00}, the increase of
chlorine coverage above 0.33 ML  leads to the
commensurate-incommensurate (C-I) phase transition via formation of
striped super heavy domain walls.

The local geometry of Cl atoms on Cu(111) has been studied
experimentally \cite{EuroPhysLett86,SS87,PRL87,SS88,Shard04} using
surface extended X-ray absorption fine structure (SEXAFS),
photoelectron diffraction and x-ray standing waves techniques
\cite{EuroPhysLett86,SS87,PRL87,SS88}. As a result, chlorine atoms
were found to be adsorbed in fcc hollow sites.

Recently, new results were obtained for the submonolayer chlorine
coverage  ($<$ 0.33 ML) on Cu(111) \cite{Cu-Cl-Chains}. Measurements
performed with low temperature scanning tunneling microscopy
(LT-STM) showed that
chlorine tends to form chain-like structures at the submonolayer stage of adsorption. In the case of single
atomic chains, chlorine atoms can alternately occupy fcc-hcp
positions corresponding to interatomic distances of 3.9 {\AA}  being
smaller than those (4.4 \AA) in the
($\sqrt{3}\times\sqrt{3}$)R30$^\circ$ structure. Indirect electronic
interaction was found  to be responsible for abnormally short Cl-Cl
interatomic distances in the chains \cite{Cu-Cl-Chains}.

First-principle calculations have been performed for the Cl/Cu(111) system in a number of works \cite{Doll_DFT,Cu_Cl_DFT,Migani_DFT,Gross}. In the first
computational work,  Doll and Harrison \cite{Doll_DFT} studied the
($\sqrt{3}\times\sqrt{3}$)R30$^\circ$ structure formed by chlorine
atoms on Cu(111) and concluded that fcc hollow  sites are
preferable. Migani et al. \cite{Migani_DFT} reported a systematic
study of all halogens on several metal surfaces. For the
Cl/Cu(111) system, the authors considered two supercells
(($\sqrt{3}\times\sqrt{3}$)R30$^\circ$ and  (3$\times$3))  and
confirmed principal conclusions of the paper by Doll and Harrison
\cite{Doll_DFT}.  Chlorine adsorption on Cu(111) was examined in the
recent study by Roman and Gross \cite{Gross}, in which
the experimentally observed monotonic increase of the work function with
chlorine coverage \cite{Goddard-Lambert} has been supported by
density-functional theory (DFT) calculations.

The most detailed work on the chlorine adsorption on the (111) plane
of copper was published by  Peljhan and Kokalj \cite{Cu_Cl_DFT}.
Authors suggested several structural models of the chlorine layer
corresponding to coverages in the range of 1/16 -- 1 ML and analyzed
their stability. One of conclusions of the paper is that only one
($\sqrt{3}\times\sqrt{3}$)R30$^\circ$ structure appears to be
stable. However, the authors cited only articles describing models
of  the uniform compression \cite{Goddard-Lambert,Jones96} and did not
take into account STM data by Andryushechkin et al.
\cite{AndryushechkinSS00,AndryushechkinUFN00} demonstrating
the formation of the uniaxially compressed domain-wall phases at
coverages exceeding 1/3 ML. Another problem that can be found in the
paper by Peljhan and Kokalj \cite{Cu_Cl_DFT} concerns the process of
the chlorine desorption from the Cu(111) surface. Using  the work of
Goddard and Lambert \cite{Goddard-Lambert} as a reference, the
authors reproduced the atomic chlorine desorption  in their
calculations. However, they did not take into account desorption in
the form of CuCl molecules reported  in Ref. \cite{Eltsov91}. Note
that the CuCl desorption was also reported by Nakakura et al.
\cite{Altman} in the case of the  Cl/Cu(100) system. In this
connection, further computational studies  are  required to clarify
the situation with desorption products.

In this paper, we present a systematic computational study of  the
molecular chlorine  adsorption on the Cu(111) surface. We tried to
describe theoretically not only all chlorine structures detected in the
STM experiments
\cite{AndryushechkinSS00,AndryushechkinUFN00,Cu-Cl-Chains} but to
examine several defect structures and test their stability. In
addition, first-principle calculations have been performed for all
stages  of the Cu(111) chlorination: dissociation of the Cl$_{2}$
molecule, adsorption of chlorine atoms, diffusion of copper and
chlorine atoms across the surface, association and desorption of
reaction products. As a result, all the contradictions between
experimental and theoretical description of the chlorine monolayer
desorption from Cu(111) have been ruled out.

\section{Theoretical Method}

The calculations were performed within the framework of DFT using
the Vienna ab-initio simulation package (VASP) \cite{VASP1,VASP2}
employing the projector augmented wave method \cite{PAW}. The
generalized  gradient approximation  with the exchange-correlation
functional form of Perdew--Burke--Ernzerhof (PBE) was applied
\cite{PBE}. The valence electron eigenfunctions are expanded in a
plane-wave basis set with an energy cutoff of 400 eV and Fermi
smearing of 0.2 eV. The equilibrium lattice parameter of bulk copper
was calculated to be 3.64~\AA, in agreement with the experimental
value of 3.61~\AA \ \cite{a_Cu_bulk}. The Cu(111) surface was
represented by a four-layer slab separated by a vacuum gap with a
thickness $\approx$ 18 {\AA}. The bottom two layers were fixed at
bulk positions while positions of atoms in the remaining copper and
chlorine layers were allowed to relax. Steps were modeled as a part
of the fifth Cu(111) layer. Reciprocal cell integrations were
performed using (7$\times$7$\times$1), (5$\times$5$\times$1), and
(2$\times$12$\times$1) Monkhorst--Pack \cite{Monkhorst-Pack} k-point
grids for the (4$\times$4), (6$\times$6), and (12$\times$2) surface
supercells, respectively. Non-spin-polarized DFT method was used for
all calculations. Spin-polarized calculations were performed for
some systems (among them single atom under the surface) to study the
impact of this effect on the results. We have examined that
non-spin-polarized full energies agree with spin-polarized ones
within the accuracy of our calculations.

The reaction barriers were evaluated using the nudged elastic band
(NEB) method \cite{NEB} implemented into VASP. Calculations of
minimal energy paths (MEP) and activation barriers on potential
energy surface (PES) were carried out with a spring constant of 5
eV/\AA$^2$. A smaller k-point grid was employed at MEP evaluation in
order to reduce the computational cost.

The adsorption energy of chlorine  on Cu(111) was calculated
according to the formula
\begin{equation}
    E_{ads}=\frac{E_{Cl/Cu(111)}-E_{Cu(111)}}{N_{Cl}} - \frac{E_{Cl_2}}{2}  ,
    \label{Eads}
\end{equation}
where $E_{Cl/Cu(111)}$, $E_{Cu(111)}$, and $E_{Cl_2}$ are the total
energy of adsorption system, clean Cu(111), and gaseous Cl$_2$ in
vacuum, respectively; $N_{Cl}$ is a number of Cl atoms adsorbed on
copper surface.

To determine the thermodynamically stable configurations of chlorine
on the Cu(111) surface, we calculate the surface free energy per
surface area $A$,
\begin{equation}
    \gamma = \frac{\Delta G - N_{Cu} \mu_{Cu} - N_{Cl} \mu_{Cl}}{A} ,
    \label{Eq_gamma}
\end{equation}
where $G$ is the Gibbs free energy, $N_{Cu}$ is a number of Cu
adatoms. Since the  chemical potential of chlorine in gas phase
($\mu_{Cl}$) is supposed usually to keep the main temperature and
pressure dependence \cite{Scheffler_gamma}, only the total energies
have been left in $G$, $\Delta G = E_{Cl/Cu(111)}-E_{Cu(111)}$. The
chemical potential of copper ($\mu_{Cu}$) is assumed to be equal to
the total energy of a copper atom in a bulk, $E_{Cu}$. Since defects
(step edges and adatoms) have been generated on surface before
chlorine adsorption, we can put $N_{Cu} =0$ and account Cu adatoms
in total energy calculations $(E_{Cu(111)})$ assuming Cu(111) as a
clean surface with defects. Now \ref{Eq_gamma} can be rewritten in
terms of $E_{ads}$:
\begin{equation}
    \gamma = \rho \theta (E_{ads} - \Delta \mu_{Cl}) ,
    \label{Eq_gamma1}
\end{equation}
where $\Delta \mu_{Cl} = \mu_{Cl} - E_{Cl_2}/2$, $\rho$ is a number
of copper atoms per surface area $A$, $\rho = 0.175$ \AA$^{-2}$,
$\theta$ is chlorine coverage in ML (1 ML is defined as a density of
surface atoms).

The desorption energy of a specie X (X = Cl, Cl$_2$, CuCl, CuCl$_2$,
Cu) from the chlorinated Cu(111) surface is defined as
\begin{equation}
    E_{des}= E_{fin} + E_{X} - E_{Cl/Cu(111)},
    \label{Edes}
\end{equation}
where $E_{fin} $, $E_{X}$, and $E_{Cl/Cu(111)}$ are the total
energies of the final surface after desorption, gaseous product X in
vacuum, and initial metal surface with adsorbate, respectively.

To put a link to the experiment, we estimated positions of
thermodesorption peaks using the values of  desorption energies
obtained in our calculations. Usually, the Polanyi--Wigner equation
is used to describe thermal desorption process:
\begin{equation}
r_{des} = k \theta^{(n)} , \ \ \ k = \nu \exp \Biggl\{ -
\frac{E_{des}}{k_B T} \Biggr\} ,  \ \,
    \label{r_des}
\end{equation}
where $r_{des}$ is the desorption rate, $n$ is the order of the
desorption kinetics, $k_B$ is the Boltzmann constant, $\nu$ is a
frequency factor ($\approx 10^{13} s^{-1}$). For the zero-order,
first-order, and second-order kinetics the dependence of
$\theta^{(n)}$ on initial adsorbate coverage ($\theta$) and
desorption time ($t$) is determined as (see, for example,
Ref.\cite{Oura})
\begin{equation}
\theta^{(0)} = \theta (1-kt/\theta) , \label{theta_0}
\end{equation}
\begin{equation}
\theta^{(1)} = \theta \exp(-kt) , \label{theta_1}
\end{equation}
\begin{equation}
\theta^{(2)} = \theta (1+k \theta t)^{-1} ,
    \label{theta_2}
\end{equation}
where $T=T_0 + \beta t $, $\beta$ is the heating rate, $T_0$ is the
initial substrate temperature. The temperature $T_m$ corresponding
to peak of the desorption rate can be found from the following
condition:
\begin{equation}
    \frac{d r_{des} }{d T} \Biggl|_{T=Tm}= 0 .
    \label{Tm}
\end{equation}

\section{Results and Discussion}

\subsection{\label{sec:diss}Dissociation of Cl$_2$ Molecule}

\begin{figure}
\centerline{\includegraphics[width=8cm]{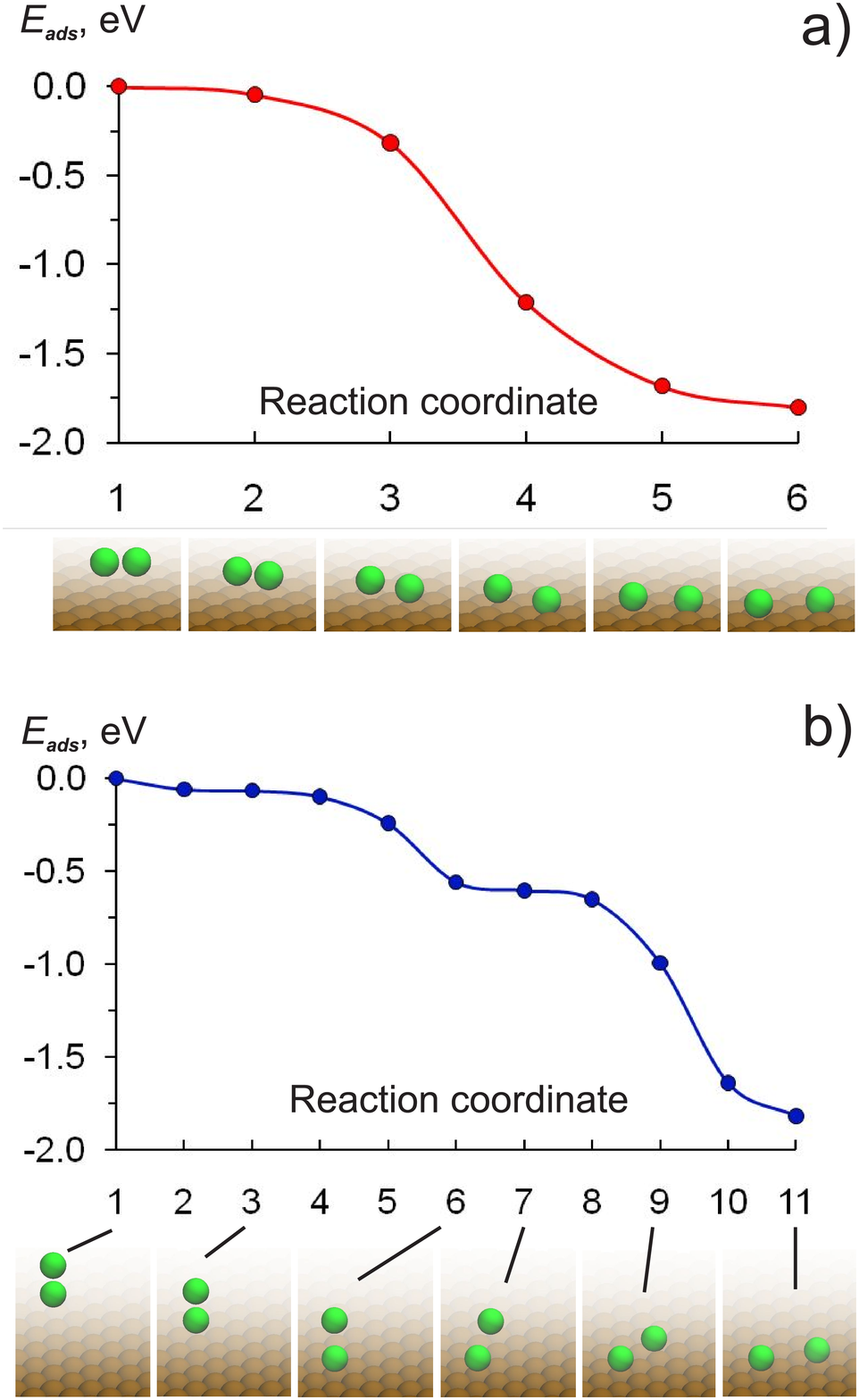}}
\caption{\label{fig_diss} Adsorption energy per Cl atom shown as a
function of the reaction coordinate for Cl$_2$ dissociation in
parallel (a) and perpendicular (b) orientations. The lines
connecting the points are guides to the eye. Images corresponded to
reaction coordinates are presented at the bottom of each graph (Cu
surface atoms are brown, Cl atoms are green). }
\end{figure}

The dissociation of molecule above metal surface may be sensitive to
its position and orientation with respect to the substrate plane
(see Ref.\cite{H2}). In this connection, we considered both parallel
and perpendicular initial orientations of the Cl$_2$ molecule with
respect to the Cu(111) plane. The minimal energy path was evaluated
between initial and final states shown in \ref{fig_diss}.

For both orientations, the initial states  were taken as minimum
height positions above the surface, in which the optimization of
coordinates would not lead to changes in the distance between
Cl$_{2}$ molecule and Cu(111). Note that we were not able to use
positions of the molecule close to the surface as the initial states
due to its fast dissociation into a pair of separate atoms during
the optimization of coordinates. The minimal distance between the
lowest Cl atom in the molecule and the plane of Cu surface atoms was
estimated to be 3.6 \AA \ and 6.3 \AA \ for parallel and
perpendicular molecule orientations, respectively. The calculated
Cl$_2$ bond length in the initial state is equal to 1.99 \AA\ that
agrees with the calculated value for the isolated chlorine molecule
and with the experimental value of 1.987 \AA \ \cite{Cl2_bond}.
According to recent DFT calculations of pair adsorption of chlorine
on the Cu(111) surface \cite{Cu-Cl-Chains}, the configuration
fcc--hcp with a separation of 3.9 {\AA} is favorable. That was a
reason to take it as the final configuration in the dissociation
process.

Calculations predict no energy barrier for the Cl$_2$ dissociation
during the adsorption in both orientations (\ref{fig_diss}). This
result is in line with previous DFT calculations of the chlorine
dissociation on Cu(111) \cite{Cu_Cl_DFT} performed only for the
parallel orientation.

For the perpendicular molecule orientation, we found a stable state
(reaction coordinate 6 in \ref{fig_diss}b) corresponding to the
adsorption energy of $-0.566$ eV per Cl atom (E$_{ads} = 0$ eV for
the initial state and E$_{ads} = -1.826$ eV for the final one).
Since MEP has the stable state, we performed two NEB-calculations:
before (reaction coordinates 1--6 in \ref{fig_diss}b) and after
(reaction coordinates 6--11 in \ref{fig_diss}b) the stable state. In
this state, the first Cl atom is placed above the second one
occupying the fcc position. During adsorption, the Cl--Cl bond
distance increases by 0.65 {\AA} and becomes equal to 2.64 {\AA}.
Note, that similar state (Cl--Cl distance of 2.34 {\AA}) has been
predicted for the case of the Cl$_2$ adsorption on Ag(111) in Ref.
\cite{Cl-Ag}.

\subsection{\label{sec:ads}Adsorption}

In this section, we present a computational study of several surface
structures formed by chlorine on the Cu(111) surface. We focused
mainly on the structures that have been reported in the experimental
STM studies
\cite{AndryushechkinSS00,AndryushechkinUFN00,Cu-Cl-Chains}. In
addition, we considered the adsorption on the step edges and near
the copper adatom.

\begin{figure}
\centerline{\includegraphics[width=7.5cm]{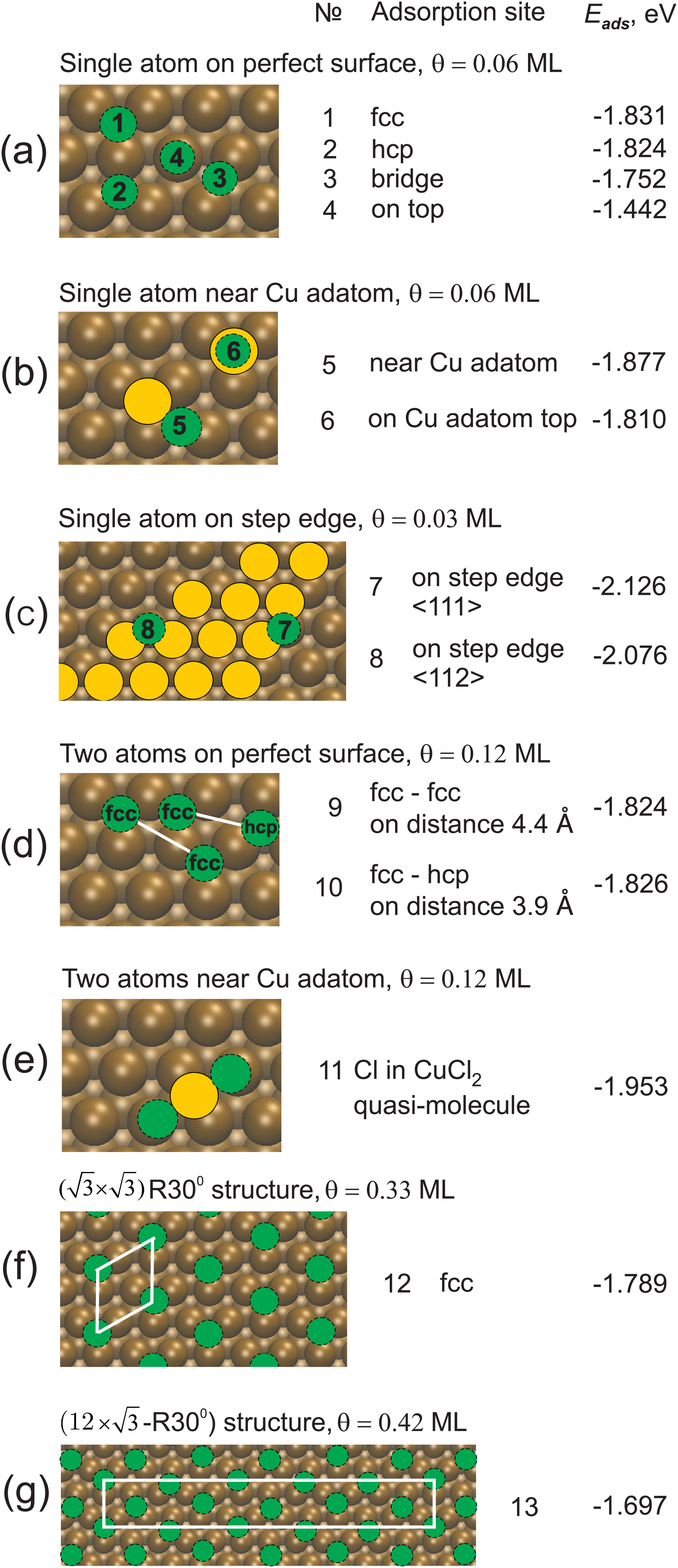}}
\caption{\label{fig_table} Atomic positions and adsorption energies
for chlorine on perfect and defective Cu(111) surface. Cu surface
atoms are brown, Cu adatoms and steps are yellow, and Cl atoms are
green. For structures at $\theta \geq 0.33$ ML unit cells are shown
by white rectangles.}
\end{figure}

At the first stage, we analyzed the adsorption of  the single
chlorine atom on Cu(111). For the (4$\times$4) unit cell, we
calculated adsorption energies for the Cl adsorption in fcc, hcp,
bridge and top sites (see \ref{fig_table}a). The fcc position was
found to be favorable (E$_{ads}=-1.831$ eV), in agreement with
results of experiments \cite{EuroPhysLett86,SS87,PRL87,SS88,Shard04}
and previous calculations \cite{Cu_Cl_DFT,Cu-Cl-Chains}.

Adsorption of the single chlorine atom was considered also near the
copper adatom placed in the fcc position on Cu(111)
(\ref{fig_table}b). According to calculations performed with the
(4$\times$4) unit cell, the position on the top of the adatom
(E$_{ads}=-1.810$ eV) is less favorable than the positions near the
adatom (E$_{ads}=-1.877$ eV) and the fcc position on the terrace.
For the optimized structure (\ref{fig_table}b, 5), the distance
between the adsorbed chlorine atom and the copper adatom is equal to
2.21~\AA. This value appears to be less  than the bond length
between the Cl atom and the nearest Cu surface atom (2.42~\AA).

\ref{fig_table}c shows  the model drawing and adsorption energies
for the single chlorine atom adsorption on the step edges. We
consider two types of steps on the (111) facet aligned parallel to
$\langle111\rangle$ and $\langle112\rangle$ directions (unit cell
(6$\times$6), coverage $\theta$ = 0.03  ML). For both types of
steps, the favorable position of the Cl atom is the threefold hollow
site between two copper atoms from the upper terrace and one copper
atom from  the lower terrace (see \ref{fig_table}c). According to
calculations, chlorine appears to be more strongly bound on the
$\langle111\rangle$ step (E$_{ads}=-2.126$ eV) than on the
$\langle112\rangle$ one (E$_{ads}=-2.076$ eV). It is noteworthy that
both values of adsorption energy  are significantly larger than
those for chlorine adsorption on terraces and near adatoms. Thus, we
can conclude that the step edges are the most attractive sites  for
chlorine adsorption on the Cu(111) surface.

\ref{fig_table}d shows the case of the pair adsorption of chlorine
atoms on Cu(111). Calculations performed within the (4$\times$4)
unit cell (coverage $\theta$ = 0.12  ML) show that the fcc--hcp
configuration (3.9 \AA) is slightly preferable in comparison with
the fcc--fcc one (4.4 \AA). This conclusion is in line with results
of the recent PBE-D2 study \cite{Cu-Cl-Chains}, in which this
difference was found to be more pronounced.

We also considered the adsorption of two chlorine atoms near the
copper adatom on Cu(111). The DFT optimized structure corresponds to
the linear geometry, in which the copper adatom and chlorine atoms
occupy bridge and on top positions, respectively (see
\ref{fig_table}e). Therefore, the adsorption of two chlorine atoms
near the surface adatom is more favorable (E$_{ads}=-1.953$~eV) than
the adsorption of two separate chlorine atoms on the Cu(111) terrace
(E$_{ads}=-1.826$~eV). Since the distances between Cl atoms and the
Cu adatom (2.16 {\AA}) appear to be less than those between Cl atoms
and the underlying copper atoms from the upper substrate layer (2.38
{\AA}), we can consider the Cl-Cu-Cl configuration as a
quasi-molecule CuCl$_2$. It is noteworthy that the arrangement of
the CuCl$_2$ quasi-molecule on the Cu(111) surface is very similar
to the case of the AuCl$_2$ quasi-molecules observed in the
Cl/Au(111) system \cite{AuCl2}.

According to STM experiments
\cite{Cu-Cl-Chains,AndryushechkinSS00,AndryushechkinUFN00},
chlorine forms on the Cu(111) surface a simple
($\sqrt{3}\times\sqrt{3}$)R30$^\circ$ structure at 0.33 ML. In this
phase, all Cl atoms occupy equivalent fcc sites as seen from
\ref{fig_table}f. The adsorption energy of chlorine in the
($\sqrt{3}\times\sqrt{3}$)R30$^\circ$ structure was found to be
equal to $-1.789$ eV.

Further chlorine dosing above 0.33 ML gives rise to the uniaxial
compression of the chlorine lattice
\cite{AndryushechkinSS00,AndryushechkinUFN00} via formation of the
striped super-heavy domain walls. At the final point of compression,
the chlorine lattice is described by the
(12$\times\sqrt{3}$-R30$^\circ$) unit cell
\cite{AndryushechkinSS00,AndryushechkinUFN00}. Results of our
calculations (optimized model and energy of adsorption) are
presented in \ref{fig_table}g. Some chlorine atoms were found to
occupy exactly fcc adsorption sites, while others --- positions
located between fcc, hcp and bridge sites. This structure  is
energetically less favorable (E$_{ads} = -1.697$ eV) in comparison
with the ($\sqrt{3}\times\sqrt{3}$)R30$^{\circ}$ structure since
only some of Cl atoms are adsorbed into the optimal fcc sites.
According to calculations, the nearest-neighbor Cl--Cl distances are
varied in the range of 3.5 \AA \ -- 4.0 \AA. Taking into account the
proximity of these values to the van der Waals diameter of chlorine
(3.6 \AA), one can conclude that further  compression of the
chlorine lattice on Cu(111) is unfavorable.

Summarizing the results shown in \ref{fig_table}, we conclude that
chlorine first adsorbs on the step edges (the maximal value of
adsorption energy). Although, there are no LT-STM images
demonstrating the presence of chlorine atoms near copper adatoms on
the Cu(111) surface, such configuration can exist at elevated
temperatures. Energies of adsorption for single and pair of chlorine
atoms on the Cu(111) terraces appear to be very close to each other.
For complete monolayers characterized by both hexagonal
($\sqrt{3}\times\sqrt{3}$)R30$^\circ$  and quasi-hexagonal
(12$\times\sqrt{3}$-R30$^\circ$) lattices, the energies of
adsorption  was found to be considerably lower than those for the
case of single and pair adsorption.

\subsection{\label{sec:gamma}Thermodynamical Stability of Surface Structures}

\begin{figure}
\centerline{\includegraphics[width=8cm]{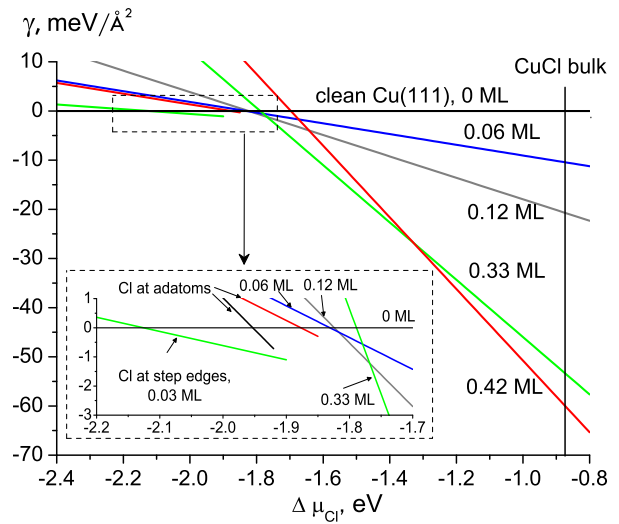}}
\caption{\label{fig_gamma} Surface free energies for Cl at Cu(111)
for low energy adsorption structures as a function of the Cl
chemical potential. Cl at adatoms and single Cl in fcc sites
correspond to 0.06 ML, CuCl$_2$ quasi-molecules and chains
(fcc--hcp) --- 0.12 ML, $(\sqrt{3}\times\sqrt{3})R30^{\circ}$
structure --- 0.33 ML, (12$\times\sqrt{3}$-R30$^{\circ}$) structure
--- 0.42 ML. The lines corresponding to Cl adsorbed at step edges
and at adatoms are interrupted because of limited capacity of these
sites. Formation energy of bulk CuCl (vertical line at $-0.83$ eV)
was calculated as $E_{CuCl}-E_{Cu}-E_{Cl_2}/2.$}
\end{figure}

To make a link between calculated adsorption structures with those
observed in the experiment, the surface free energy ($\gamma$) was
evaluated as a function of the Cl chemical potential ($\Delta \mu_{Cl}$)
by \ref{Eq_gamma1}. The thermodynamically most stable structure at
given $\Delta \mu_{Cl}$  corresponds to the lowest energy line in
\ref{fig_gamma}. The horizontal line at $\gamma = 0$ represents the
clean surface.

The clean copper surface is the most stable at the chemical potential
smaller than the lowest adsorption energy for chlorine on Cu(111),
$\Delta \mu_{Cl} < -2.13 $ eV. Chlorine at step edges is the most
stable configuration from $\Delta \mu_{Cl} = -2.13$ eV and until
these sites become completely filled.

On terraces, Cl atoms in CuCl$_2$ quasi-molecules and near Cu
adatoms are most thermodynamically stable at $\Delta \mu_{Cl} >
-1.95 $~eV and $\Delta \mu_{Cl} > -1.88 $~eV, respectively, and till
adatoms are present on the surface. Single Cl atoms are stable in a
tiny range $-1.83$ eV $ < \Delta \mu_{Cl} < -1.82 $ eV. At $-1.82$
eV $ < \Delta \mu_{Cl} < -1.77 $ eV, chains are thermodynamically
preferred.

The $(\sqrt{3}\times\sqrt{3})R30^{\circ}$ structure is stable in the
 range of $-1.77$ eV $ < \Delta \mu_{Cl} < -1.33 $ eV.
Increase of the chlorine chemical potential above $ \Delta \mu_{Cl}
= -1.33 $ eV leads to the (12$\times\sqrt{3}$-R30$^{\circ}$)
structure, which is stable up to $\Delta \mu_{Cl} = -0.83$ eV, i.e.
to the CuCl bulk formation energy.

\subsection{\label{sec:diff}Diffusion}

Diffusion of chlorine and copper atoms on the Cu(111) surface  may
effect on the kinetics of the adsorbate layer transformation and on the
rearrangement of copper  atoms on terraces and steps edges. In the
present work, we used the NEB calculations to study the mobility of
Cl and Cu atoms on both perfect and defective Cu(111) surfaces.

According to our calculations, the minimal energy paths (MEP) passes
through the bridge site  playing a role of the transition state. The
energy barrier for the fcc--hcp migration of the Cl atom is equal to
the difference between adsorption energies of chlorine in fcc and
bridge sites. The height of the energy barrier was evaluated as 0.08
eV, in agreement with earlier computations \cite{Cu_Cl_DFT}, in
which this value was reported to be less than 0.1 eV.

\begin{figure}
\centerline{\includegraphics[width=8cm]{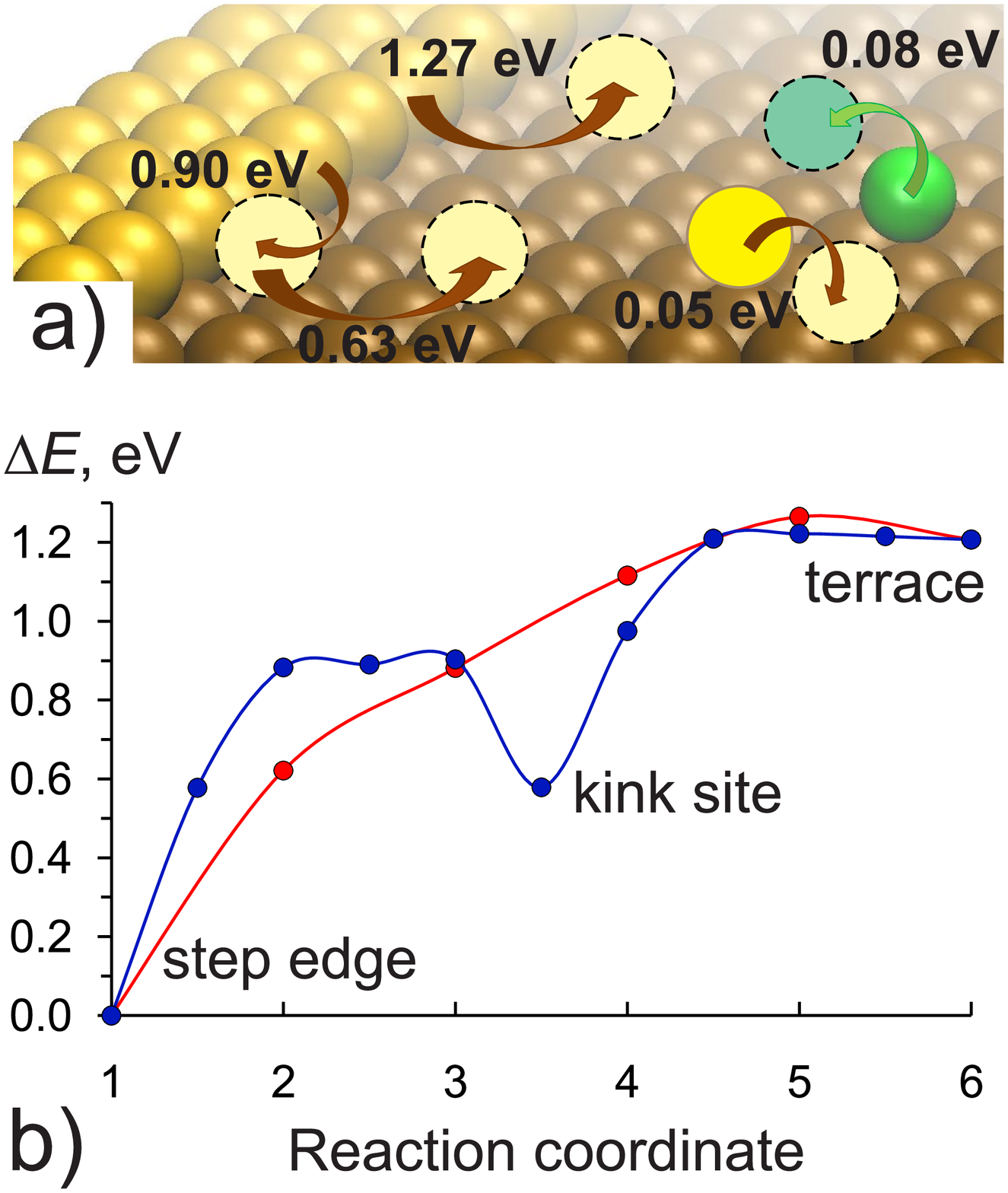}}
\caption{\label{fig_diff}a) Scheme of considered diffusion paths on
the Cu(111) surface. Cu surface atoms are brown, Cu adatoms and
steps are yellow, and Cl atoms are green. b) Activation energy as a
function of reaction coordinate for Cu diffusion from step edge to
terrace directly (red curve) and via kink site (blue curve).}
\end{figure}

The height of the energy barrier for the fcc--hcp copper adatom
hopping  on the terrace was found to be equal to 0.05 eV (taking into
account that MEP is the same as for the Cl atom diffusion). The
estimated barrier height for the direct Cu atom movement from the
step edge to the terrace is equal to 1.27 eV (\ref{fig_diff}). In
the other scenario, the copper atom can migrate to the terrace through a
kink site of the step. The transition of Cu from the step edge to the kink site
requires the activation energy of 0.90~eV, and from kink site to the
terrace --- 0.63 eV (\ref{fig_diff}). It means that at energies
higher than 0.90 eV, copper atoms can migrate along step edges inducing a change in their shape.

\subsection{\label{sec:des}Association \& Desorption}

To investigate the desorption of chlorine from the Cu(111) surface,
we calculated   activation energies for probable desorption
products: Cl, Cl$_2$, CuCl, CuCl$_2$, and Cu. According to the
transition state theory, MEP for the dissociative adsorption is the
same as for the associative desorption.

Since the Cl$_2$ molecule dissociates above Cu(111) without any
activation barrier, the associative desorption of molecular chlorine
also has no additional barrier. Therefore, the energy barrier of the
associative Cl$_2$ desorption is determined only by the energy
difference between the initial and finite states and could be
calculated by \ref{Edes}.

We carried out the NEB-calculations of CuCl and CuCl$_2$ molecules
adsorption on Cu(111). As initial states for MEP calculation, we
considered the perpendicular orientation of these molecules with
respect to the substrate plane. The calculated Cl-Cu distance in the
initial state of the CuCl molecule is equal to 2.04~\AA. In the
initial configuration of the CuCl$_2$ molecule, the Cl-Cu  and Cl-Cl
distances were found to be  2.05~\AA \  and 4.11 {\AA},
respectively. Note that structural parameters of the both molecules
in the initial state above the Cu(111) surface agrees well with the
calculated values for the isolated molecules in vacuum. In the final
states, the molecules stay on the surface (see \ref{fig_table}b (5)
and \ref{fig_table}e) and the bonds are slightly stretched.
According to our calculations, CuCl and CuCl$_2$ molecules adsorb on
Cu(111) without any activation barrier. The same conclusion was made
in Ref. \cite{CuCl_abstraction}, wherein the desorption of the
on-top adsorbed chlorine atom together with the underlying copper
atom from the substrate has been considered.

\ref{tab:table_des} summarizes the desorption energies for probable
desorption products from the chlorinated Cu(111) surface.
Calculations have been performed for both perfect and defective
surfaces. The structural model of defective surface contains two
types of steps (aligned  parallel to $\langle$111$\rangle$ or
$\langle$112$\rangle$ directions) with several adsorption sites.
However, we present here only the lowest energy configuration.

Next, we analyze desorption processes for each specie depending on
the chlorine coverage.

\begin{table*}
  \caption{Desorption energies (in eV) calculated for probable desorption products from the chlorinated Cu(111) surface}
  \label{tab:table_des}
  \begin{tabular}{cccc}
    \hline
$\theta$, ML&  \textrm{desorption sites}& \textrm{specie}& $E_{des}$ \\
%\colrule
\hline
0    & adatom & Cu & 2.69 \\
     & step edge & Cu  & 3.42 \\
     & terrace & Cu  & 4.47 \\
 \\
0.03 & step edge & CuCl & 2.92 \\
     & step edge & Cl & 3.58 \\
     & step edge & CuCl$_2$ & 3.66 \\
     & step edge & Cl$_2$ & 4.14 \\
 \\
0.06 -- & Cl with Cu adatom & CuCl & 2.18\\
0.12 & CuCl$_2$ quasi-molecule & CuCl$_2$ & 2.60\\
     & fcc site & Cl &  3.33 \\
     & chain (fcc--hcp) & Cl$_2$ & 3.64 \\
     & Cl from fcc and Cu from terrace & CuCl & 3.71 \\
\\
0.33 & step edge & CuCl & 3.02 \\
     & ($\sqrt{3}\times\sqrt{3}$)R30$^\circ$ & Cl  & 3.30 \\
     & step edge & CuCl$_2$ & 3.41 \\
     & step edge & Cl & 3.57 \\
     & ($\sqrt{3}\times\sqrt{3}$)R30$^\circ$  & Cl$_2$  & 3.59 \\
     & Cl from ($\sqrt{3}\times\sqrt{3}$)R30$^\circ$ and Cu from terrace & CuCl   & 3.66 \\
 \\
0.42 & step edge & CuCl & 2.59 \\
     & (12$\times\sqrt{3}$-R30$^{\circ}$) & Cl & 3.03\\
     & Cl from (12$\times\sqrt{3}$-R30$^{\circ}$) and Cu from terrace& CuCl & 3.14\\
     & step edge & Cl & 3.20 \\
     & (12$\times\sqrt{3}$-R30$^{\circ}$) & Cl$_2$ & 3.29\\
     & step edge & CuCl$_2$ & 3.37 \\
    \hline
  \end{tabular}
\end{table*}

\paragraph{Cl desorption.}
As expected, the energy required to break bonds between chlorine and
surface copper atoms decreases as chlorine atoms start to occupy
less favorable adsorption sites (see \ref{tab:table_des}). Moreover,
at any coverage the atomic desorption of chlorine from step edges is
less favorable than that from the perfect surface.

\paragraph{Cl$_2$ desorption.}
The Cl$_2$ desorption demonstrates qualitatively very similar
behavior as the atomic Cl desorption. According to our calculations,
the desorption of the  Cl$_2$ molecule from the
(12$\times\sqrt{3}$-R30$^{\circ}$) structure (3.29 eV) is easier
than that of  other structures
(($\sqrt{3}\times\sqrt{3}$)R30$^{\circ}$  --- 3.59 eV, two
individual atoms in fcc--hcp sites --- 3.64 eV, step edge --- 4.14
eV). Thus, in the desorption process, chlorine is released first from less favorable adsorption sites populated only at high coverage (0.42 ML).

\paragraph{CuCl desorption.}
The easiest way for the CuCl desorption is to remove the  Cl  atom
together with the Cu adatom (2.18 eV). The activation barrier for
the CuCl desorption from the step edges appears to be higher (2.59
eV for the coverage of 0.42 ML, 3.02 eV for 0.33 ML, and 2.92 eV for
0.03 ML). The desorption of the CuCl molecule from the Cu(111)
terrace is less favorable. Indeed, the activation barrier of the
CuCl desorption is equal to 3.14 eV for the coverage of 0.42 ML,
3.66 eV for the coverage of 0.33 ML, and 3.71 eV for the coverage of
0.06 ML. The CuCl desorption energy is defined mainly by the number
of the nearest Cu atoms: the less nearest-neighbors Cu are around
the CuCl molecule, the less energy is required to break Cu--Cu
bonds.

\paragraph{CuCl$_2$ desorption.}
Using analogy with the CuCl desorption, we believe that the easiest
way to desorb CuCl$_2$ is to remove two chlorine atom together with
the copper adatom or with the copper atom from the step edge. That
is why, we did not consider the possible extraction of the copper
atom from the upper Cu(111) plane. According to our calculations,
the CuCl$_2$ desorption barrier appears to be higher than that for
the desorption of the CuCl molecule (see \ref{tab:table_des}).

\paragraph{Cu desorption.}
According to our calculations, the activation energy of the
desorption of copper atoms from the Cu(111) surface increases in the
series: adatoms, step edges, and first surface layer (see
\ref{tab:table_des} at zero chlorine coverage). As in the case of
the CuCl desorption, the less number of copper atoms in the first
coordination sphere, the easier Cu atom can be removed. Our
calculations show that copper sublimation starts at 2.69 eV (adatoms
removal). This energy considerably exceeds the activation energy of
copper atoms diffusion from step edges to the terrace (1.27 eV).
Therefore, the desorption mechanism via adatoms removal seems to be
plausible.

\begin{figure}
\centerline{\includegraphics[width=8cm]{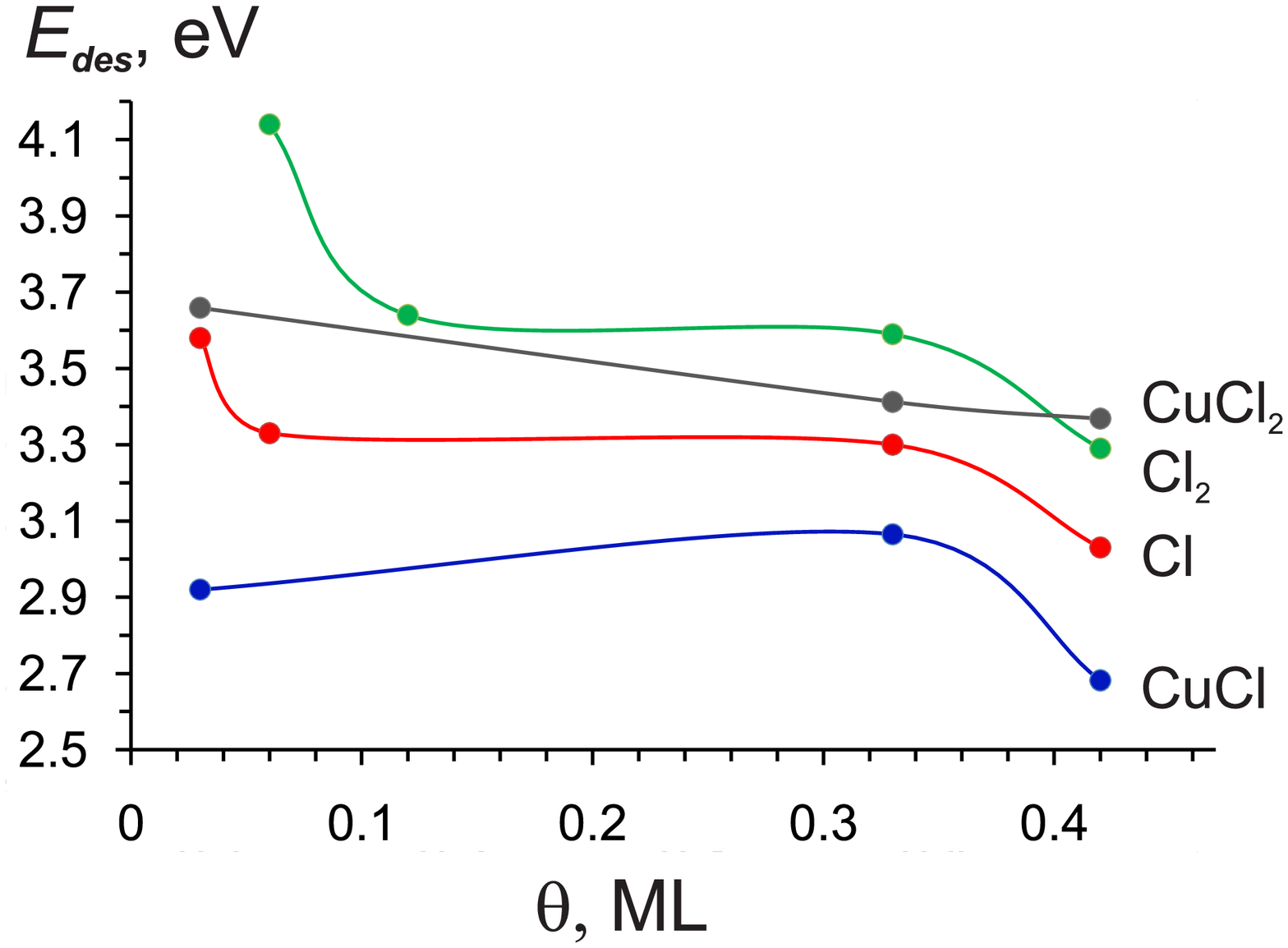}}
\caption{\label{fig_theta} Minimal desorption energies for Cl,
Cl$_2$, CuCl, and CuCl$_2$ shown as functions of chlorine coverage.
The lines connecting the points are guides to the eye.}
\end{figure}

Thus, analysis of our computational results obtained for desorption
of  Cl, Cl$_2$, CuCl, and CuCl$_2$, shows that copper (I) chloride
has the lowest barrier height at all coverages up to 0.42 ML. For
the  better recognition, we present our data on the plot showing the
dependence of the desorption energy on the chlorine coverage
(\ref{fig_theta}). In \ref{fig_theta}, we presented only the minimal
desorption energy for each specie for the given coverage. Since
there is no experimental evidence of Cu adatom presence at chlorine
coverage > 1/3 ML, corresponded to the starting coverage in
desorption experiments, desorption of CuCl and CuCl$_2$ molecules
from adatoms is not shown in \ref{fig_theta}.

Now we can describe the desorption mechanism as following. The
desorption of chlorine takes place in the form of CuCl molecules at
any coverage. Desorption preferably occurs from the step edges
(minimal barrier for desorption). Since the chlorine adsorption on
the step edges is energetically more favorable than the adsorption
on the terrace, the chlorine population at the step edges is
recovered continuously as a result of fast (energy barrier is equal
to 0.08 eV) diffusion of chlorine atoms from the terrace.

These conclusions are in line with thermodesorption data published
in the work by Eltsov et al. \cite{Eltsov91}, in which the
desorption of chlorine monolayer was found to take place in the form
of CuCl molecules in the temperature range of 700--850 K (peak
maximum $\approx$800 K). Taking into account that in our
calculations the CuCl desorption corresponds to minimal activation
energy (2.59 -- 2.92 eV at $\theta$= (0.42 -- 0.03) ML) with respect
to other probable desorption products (Cl, Cl$_{2}$, CuCl$_2$), we
can make a rough estimation of the position of the CuCl desorption
peak. Using experimental conditions taken from work by Eltsov et al.
\cite{Eltsov91} ($\theta=0.42$ ML, $\beta=2$ K/s, $T_0=160$ K), we
obtained values 840 K and 859 K for first-order (\ref{theta_1}) and
second-order (\ref{theta_2}) desorption for $E_{des} = 2.59$ eV,
respectively. Thus, a good agreement with experimental data has been
achieved.

In their work, Goddard and Lambert \cite{Goddard-Lambert} reported
atomic chlorine as the only desorption product. The reported
activation energy of desorption was estimated as 236 kJ/mol (2.45
eV) that is close to our data for the CuCl desorption from steps but
not with atomic chlorine desorption. In our belief, the
contradiction can be solved if assume that in the experiments by
Goddard and Lambert \cite{Goddard-Lambert} the sensitivity of the
mass-spectrometer was likely not enough to detect CuCl species.

\section{\label{sec:conc}Conclusions}

We have computationally studied the reaction of molecular chlorine
with both clean and defective Cu(111) surfaces. It was found that the Cu(111) surface has a strong catalytic effect
on the Cl$_2$ dissociation leading to the elimination of the activation
barrier at both perpendicular and parallel orientations of the
molecule axis. Copper and chlorine atoms have negligibly small
diffusion barriers on Cu(111) allowing the transformation of the surface
phases at $E > 0.1$ eV. Step edges are more stable and can be
modified at higher activation energies ($E > 0.9$ eV).

The most energetically favorable adsorption sites for chlorine are
step edges. In this positions, the Cl atom is attached to two Cu
atoms from the upper terrace and occupies position directly above
one Cu atom from the low terrace.

Chlorine desorption from the Cu(111) surface takes place in the form
of CuCl molecules. The CuCl species desorb first from step edges (at
$\theta < 0.42$ ML). After each event of the CuCl desorption, the
population of chlorine atoms at the step edges  is recovered
continuously as a result of fast diffusion of chlorine atoms from
the terrace.

\section{Acknowledgements}  Authors are grateful to K.N. Eltsov for ideas and
helpful discussions. The reported study was supported by RFBR,
research project 15-02-99607\_a. We thank the Joint Supercomputer
Center of RAS and the Supercomputing Center of Lomonosov Moscow
State University \cite{MGU} for the use of their computational
facilities.
%%%%%%%%%%%%%%%%%%%%%%%%%%%%%%%%%%%%%%%%%%%%%%%%%%%%%%%%%%%%%%%%%%%%%
%% The same is true for Supporting Information, which should use the
%% \suppinfo macro.
%%%%%%%%%%%%%%%%%%%%%%%%%%%%%%%%%%%%%%%%%%%%%%%%%%%%%%%%%%%%%%%%%%%%%
%\suppinfo

%%%%%%%%%%%%%%%%%%%%%%%%%%%%%%%%%%%%%%%%%%%%%%%%%%%%%%%%%%%%%%%%%%%%%
%% The appropriate \bibliography command should be placed here.
%% Notice that the class file automatically sets \bibliographystyle
%% and also names the section correctly.
%%%%%%%%%%%%%%%%%%%%%%%%%%%%%%%%%%%%%%%%%%%%%%%%%%%%%%%%%%%%%%%%%%%%%
%\bibliography{achemso}

\end{document}